\documentclass[preprint,showpacs,showkeys,aps,prd,amsfonts,eqsecnum,nofootinbib,floatfix]{revtex4}
\usepackage{graphicx,epsfig}
\usepackage[dvips]{color}
\usepackage{amsmath, amssymb, graphics}

\begin{document}
 \def\la{\langle}
 \def\ra{\rangle}

\preprint{CUMQ/HEP 148}
%
%
% Title of the paper
\title{Production of Single Heavy Charged Leptons at a Linear Collider}
\author{Erin De Pree}\email[]{ekdepr@wm.edu}
\author{Marc Sher}\email[]{mtsher@wm.edu}
\affiliation{Particle Theory Group, Department of Physics, College of William and Mary, Williamsburg, VA 23187, 
USA}
\author{Ismail Turan}\email[]{ituran@physics.concordia.ca}
\affiliation{Department of Physics, Concordia University, 7141
Sherbrooke Street West, Montreal, Quebec, CANADA H4B 1R6}
\date{\today}

\begin{abstract}
    A sequential fourth generation of quarks and leptons is allowed by 
    precision electroweak constraints if the mass splitting between the heavy 
    quarks is between 50 and 80 GeV.  Although heavy quarks can be easily 
    detected at the LHC, it is very difficult to detect a sequential heavy 
    charged lepton, $L$, due to large backgrounds.  Should the $L$ mass be 
    above $250$ GeV, it can not be pair-produced at a 500 GeV ILC.  We 
    calculate the cross section for the one-loop process $e^+e^-\rightarrow L\tau$.
    Although the cross section is small, it may be detectable.
      We also consider contributions 
    from the two Higgs doublet model and  the 
    Randall-Sundrum model, in which case the cross section can 
    be substantially higher.
\end{abstract}
%\vskip -2.5cm
\maketitle

\section{Introduction}

One of the simplest extensions of the Standard Model is the addition 
of a new, sequential generation of fermions.   Interest in this fourth 
generation has waxed and waned over the years.   After precision 
measurements of the Z width showed that there are precisely three 
weakly interacting neutrinos \cite{three}, it became clear that the neutrino mass 
of a fourth generation would have to exceed $45$ GeV 
and interest faded.

During the 90's there was intensive study of the phenomenology of 
additional quarks and 
leptons which were {\it not} sequential \cite{fhs,later}.   Many grand unified 
theories have additional fermions, such as vectorlike isosinglet 
quarks and leptons, additional vectorlike states arise in 
gauge-mediated supersymmetry breaking models, and many additional 
models contain mirrorlike fermions.   These models are still of 
interest, but they do not require sequential fermions (although they can 
accommodate them).

Interest in a sequential fourth generation faded further with studies 
of precision electroweak constraints.  The recent Particle Data 
Group analysis claimed that ``An extra generation of ordinary
fermions is excluded at the 99.999\% CL on the basis of the S
parameter alone''\cite{three}.   However, this analysis assumes a
mass-degenerate fourth generation.   Since one of the most striking 
features of the mass spectrum of the first three generations is the 
wide range of masses, such an assumption may not be justifiable.

Analyses of the effects of a non-degenerate sequential fourth 
generation originally focused on the case where the neutrino mass was of 
$O(50)$ GeV  \cite{he,maltoni,novikov} and used 2001 electroweak 
data.  A comprehensive analysis of the current status of precision 
electroweak fits and a fourth generation was recently carried out by 
Kribs, Plehn, Spannowsky and Tait\cite{kribs}.   They noted that the 
constraints on the oblique parameters from combined electroweak data 
have been  determined by both the LEP Electroweak Working Group \cite{lep} 
and the Particle Data Group \cite{three}.   The two groups used 
somewhat different datasets and differ by roughly one standard 
deviation (see Ref. \cite{kribs} for a detailed discussion of the 
differences).  Kribs et al. used the LEP Electroweak Working Group 
results, and found that a substantial region of fourth-generation 
parameter space is in agreement with all experimental constraints.  
In this region of parameter space, the mass splitting between the $U$ 
and $D$ quarks is between $50$ and $80$ GeV.   Bounds on the mass
splitting between the charged lepton, $L$, and the neutrino, $N$, are 
less constrained since one considers both Dirac and Majorana 
neutrino masses \cite{gates}.

Thus, we see that precision electroweak data do not exclude a 
sequential fourth generation.   U and D quark production at the 
Large Hadron Collider (LHC) will be relatively easy to detect.
However, the heavy charged lepton, $L$, will be substantially more 
difficult to detect, primarily due to large backgrounds.  
Early LHC and SSC studies \cite{ehlq,willenbrock} made the assumption 
of a massless fourth generation neutrino, and still concluded that 
detecting a heavy lepton with a mass greater than $250$ GeV might not be 
possible.  Calculations of heavy lepton production (with a heavy 
neutrino) exist \cite{ng}, but do not include any discussion of 
signatures or backgrounds.  Therefore, it is likely that a heavy charged 
lepton with a mass greater than $250$ GeV will not be detected at the 
LHC.

At the International Linear Collider (ILC), heavy leptons can be easily 
 produced and detected up to the kinematic limit.  However, the 
initial stage of the ILC will probably be at a center of mass energy 
of $500$ GeV, in which case pair production of heavy leptons with 
masses above $250$ GeV will not be possible.  The only possible 
production mechanism would be through single $L$ production, in 
association with a lighter Standard Model charged lepton.  Since mixing between $L$ 
and $\mu$ or $e$ is expected to be small, we will focus on the process 
$e^{+}e^{-}\rightarrow L\tau$, which can occur through a nonzero $\theta_{34}$ 
mixing angle.   Although single production of heavy charged leptons 
has been studied before \cite{fhs,later}, all of these studies 
considered vectorlike or mirrorlike leptons, and we know of no 
calculations of this process with a heavy neutrino at a linear 
collider.  An analysis of sequential heavy charged leptons in 
$Z$-decays \cite{illana} ignored the mass of the heavy 
neutrino.

In the next section, we present the relevant diagrams in the sequential standard model and calculate the cross section as a function of the $N$ and 
$L$ masses.   Section 3 considers the cross section in 
the two Higgs doublet model and in the Randall-Sundrum model.  Finally, in Section 4, we discuss 
detection possibilities and present our conclusions.

\section{Sequential Standard Model}
    
A single charged heavy lepton can only be produced if the fourth 
generation mixes with the lighter generations.  Bounds on the mixing 
angle $\theta_{34}$ arise from 
observation of universality in $\tau$ decays; a nonzero mixing angle 
would multiple the rate by $\cos^{2}\theta_{34}$.  This was analyzed 
by Swain and Taylor \cite{swain} who found a model-independent bound 
of $\sin^{2}\theta_{34} < 0.007$.   This is a particularly 
interesting value.  If one has a Fritzsch-type $2\times 2$ mass 
matrix for the $\tau$ and $L$, one might expect $\sin^{2}\theta_{34}$ 
to be approximately $m_{\tau}/m_{L}$, which gives $0.007$ for an $L$ 
mass of $250$ GeV.  We will assume this value of the mixing angle in 
our numerical results, and can easily scale the cross section for 
smaller mixing angles.
\begin{figure}[bth]
%\vspace*{-3.1in}
        \centerline{\epsfxsize 6.4in {\epsfbox{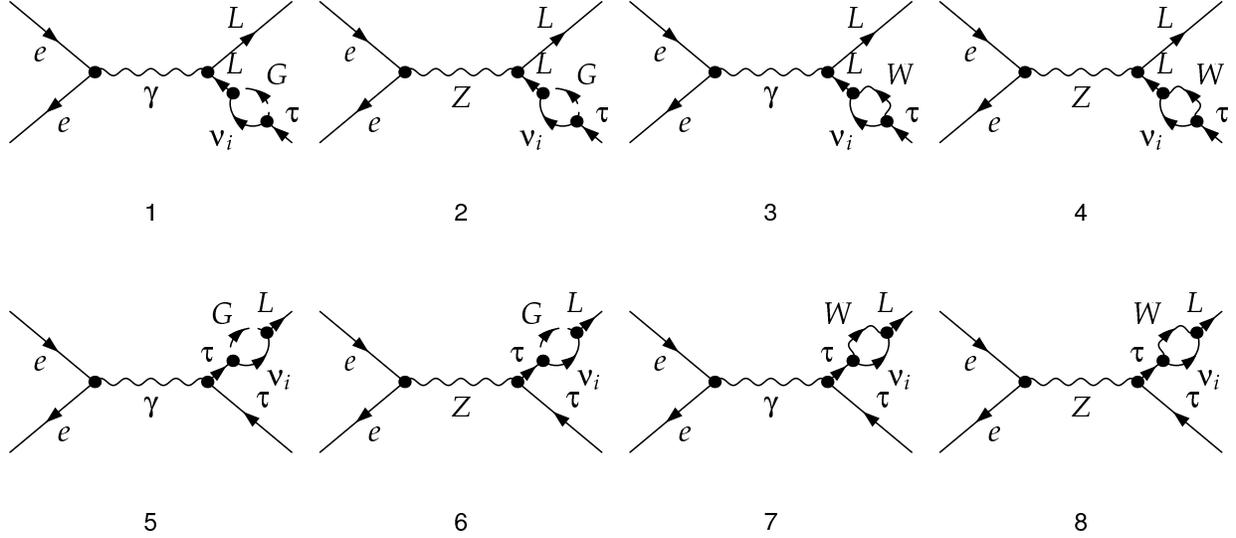}}}
%\vspace*{-2.35in}
\caption{The leading order self energy contributions to the $e^+e^-\to L\bar{\tau}$ process in the sequential SM. The Feynman gauge is assumed and the light electron-Higgs couplings are neglected.}
\label{fig:SMself}
\end{figure}
\begin{figure}[bth]
\vspace*{-0.1in}
        \centerline{\epsfxsize 6.4in {\epsfbox{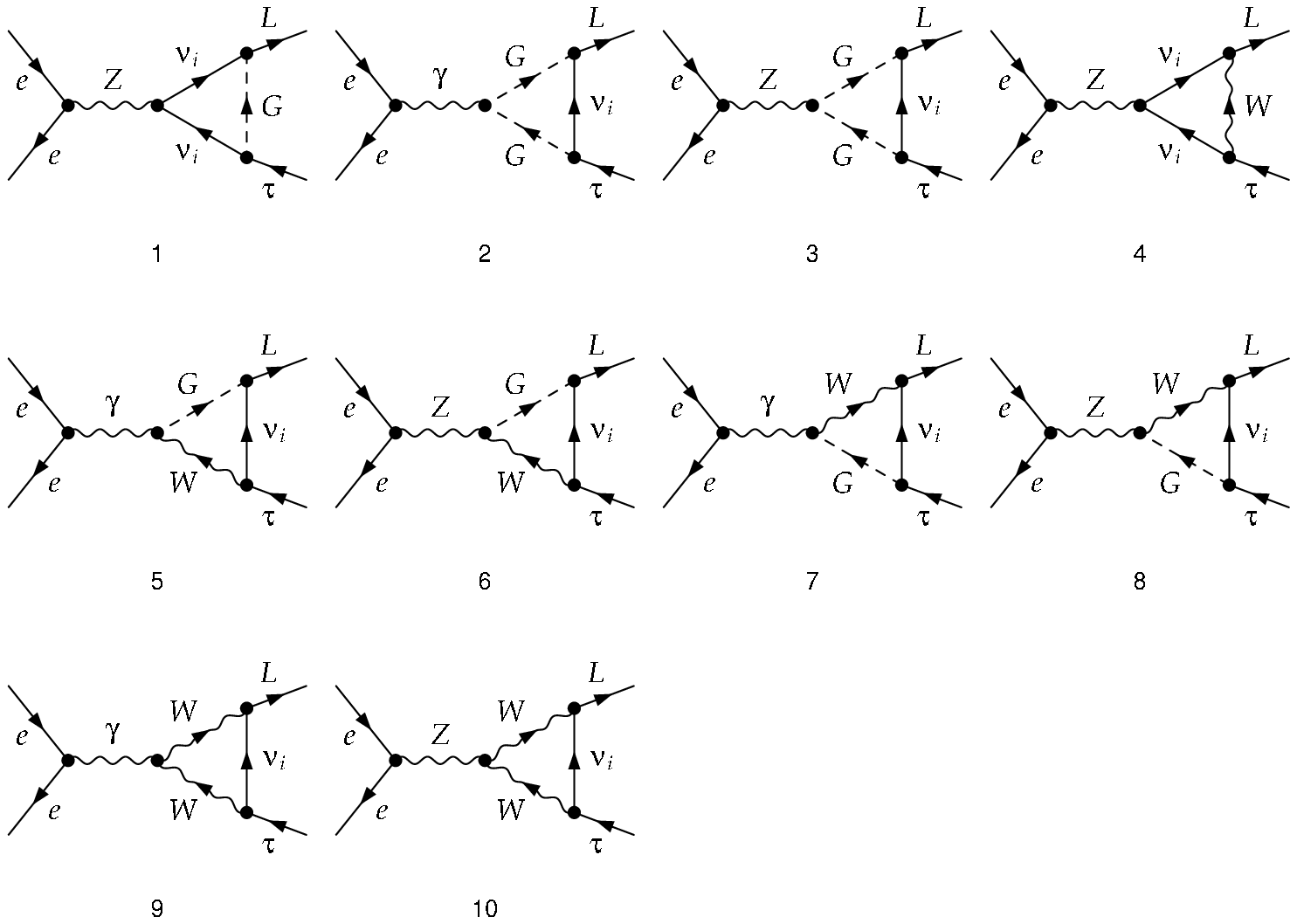}}}
%\vspace*{-2.35in}
\caption{The leading order vertex contributions to the $e^+e^-\to L\bar{\tau}$ process in the sequential SM.The `t-Hooft-Feynman gauge is assumed and the light electron-Higgs couplings are neglected.}
\label{fig:SMvertex}
\end{figure}
%\clearpage
\begin{figure}[bth]
%\vspace*{-0.6in}
%\vspace*{-0.4 cm}
        \centerline{\epsfxsize 2in {\epsfbox{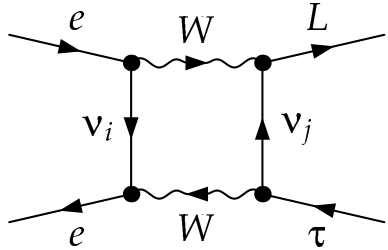}}}
%\vspace*{-2.35in}
\caption{The leading order box contributions to the $e^+e^-\to L\bar{\tau}$ process in the sequential SM. The `t-Hooft-Feynman gauge is assumed and the light electron-Higgs couplings are neglected.}
\label{fig:SMbox}
\end{figure}

The diagrams are listed in Figures 1-3  and grouped as the self energy, vertex and box type contributions, respectively.   We use the `t-Hooft-Feynman 
gauge throughout, and thus charged Goldstone bosons, $G$, must be included. 
Note that the electron-Higgs couplings are neglected due to small Yukawa couplings.
The internal neutrino lines get a contribution from each of the four 
neutrinos, and thus each diagram is proportional to 
$V_{4i}^{*}V_{i3}$.  When summing over the four neutrinos, parts of 
the matrix elements that are independent of the neutrino mass will 
cancel by unitarity of the 4-D CKM-like matrix.   This causes the 
ultraviolet divergences to cancel in the sum over neutrinos.
\begin{figure}[bth]
%\vspace*{-3.1in}
        \centerline{\hspace{-3cm}\epsfxsize 5.5in {\epsfbox{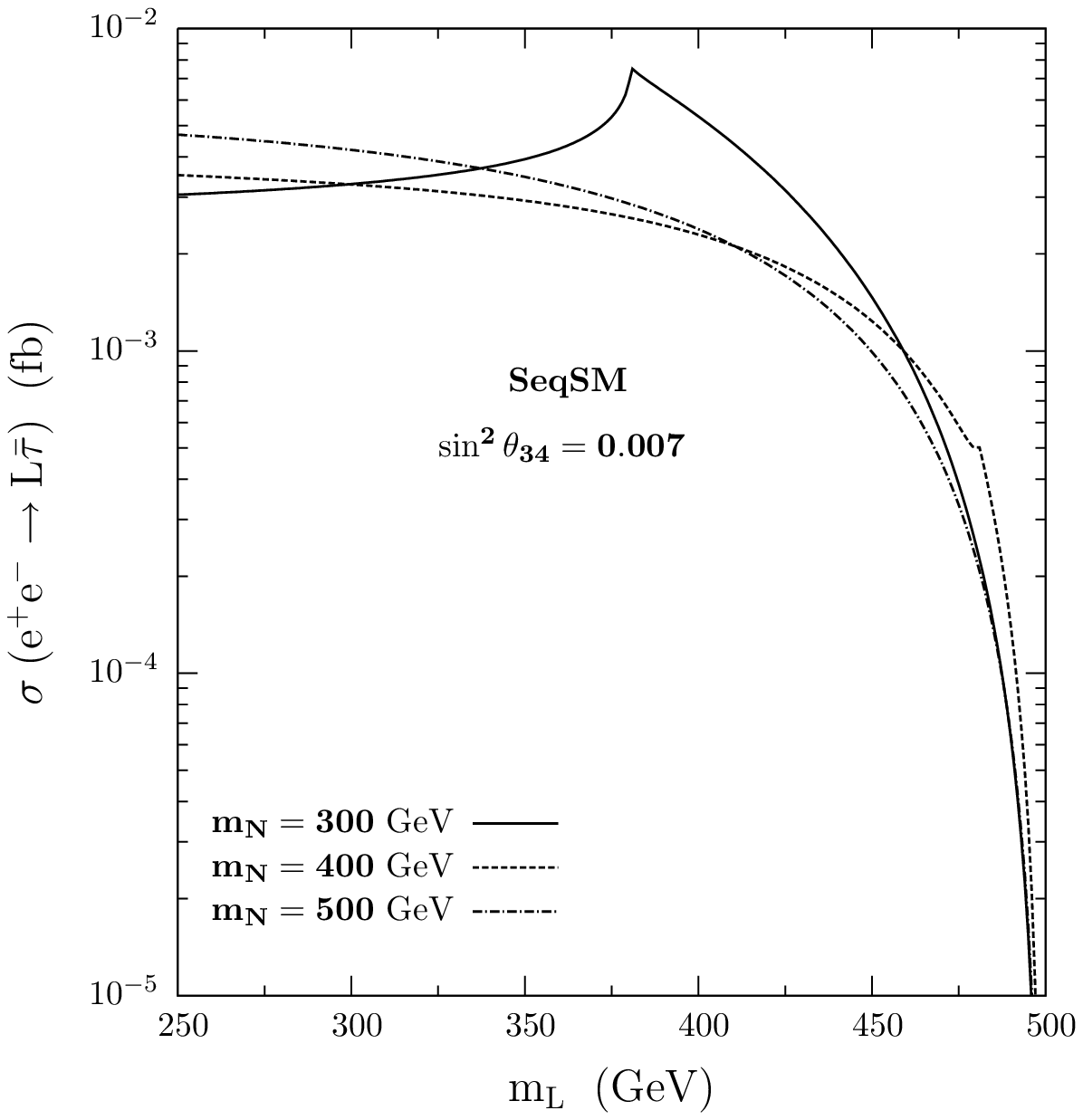}}}
\vskip -0.1in
\caption{The total cross section of $e^+e^-\to L\bar{\tau}$ as a function of the heavy lepton mass $m_L$ for $\rm \sqrt{s} = 500$ GeV and various heavy neutrino  masses in an unpolarized electron-positron beam within the Sequential SM framework.}
\label{fig:sigma300}
\end{figure}

The calculation of the cross section is performed by using 
the  {\tt FeynArts}, {\tt FormCalc}, and {\tt LoopTools} packages 
\cite{FFL}.  We first patched the SM and 2HDM model files of the {\tt FeynArts} package by 
introducing the fourth generation leptons and their interactions.  Then, 
the numerical analysis was carried out in Fortran with the help of {\tt FormCalc} and {\tt LoopTools}. 
The cancellation of the ultraviolet and infrared divergences has been checked 
numerically and the expected cancellation was confirmed. In addition, as a separate check the expected null result 
for the cross section due to unitarity of the mixing matrix $V_{ij}$ 
was also tested numerically by setting the heavy neutrino mass $m_N$ 
to zero.   Note that the same technique is applied for the calculation in the 2HDM, presented in the next section.

The results are plotted in Figure 4 for neutrino masses of $300, 400$ 
and $500$ GeV.  
 We see that, for $\sin^{2}\theta_{34} = 0.007$, cross 
sections of the order of a few attobarns can be expected.   We can 
also show that as the 
neutrino mass increases, the cross section grows rapidly, reaching 500 
attobarns at $M_{N}=2000$ GeV.   This is not surprising since the 
theory is chiral.   Of course, the cross section scales with 
$V_{34}^2$; the value we have chosen is the maximum allowed 
from the analysis of Swain and Taylor.

The structure of the curves in Figure 4 can be easily understood.   Since 
the theory is chiral, one expects the cross section to increase 
as the mass of the heavy neutrino $m_N$ increases. However, as seen from 
Figure 4 that this is not necessarily true for neutrinos in the 
$300-400$ GeV mass range. One can understand, 
for example, why the the curve for $m_N=300$ GeV crosses and becomes bigger than the one for $m_N=400$ GeV and similar behavior occurs 
between the $m_N=400$ GeV and $m_N=500$ GeV curves. This is simply due 
to the fact that both the $W$ boson and the heavy neutrino $N$ go 
on-shell in the loop if  the condition $m_L\ge m_W + m_N$ is kinematically 
satisfied. When $m_L$ is large enough to produce the $W$ and 
$N$ on-shell, the loop integrals develop imaginary parts, which can be calculated by using the Cutkosky rules, and results in enhancement of 
the cross section. One can calculate this by cutting through the $W$ boson-$\nu_i$ propagators (for $i=4$) at the heavy lepton's leg 
in Figures 1-3. Thus, for example, the peak due this enhancement for the $m_N=300$ GeV curve occurs at around $m_N + m_W$ and it shifts to the 
right for the $m_N=400$ GeV curve.

%\begin{figure}[t]
    %\centering
    %\includegraphics[height=7cm, viewport=180 180 500 550]{fig2.eps}
    %
    %\caption{Cross section for $e^+e^-\rightarrow L\tau$ for an $N$ mass of 300 GeV and a mixing 
    %   angle equal to its largest allowed value of $\sin^2\theta_{34}=0.007$}
    %\label{fig:figtwo}
%\end{figure}

Are these small cross sections detectable?  With an integrated 
luminosity of an inverse attobarn, expected at the ILC's full luminosity for a 
couple of years, one expects a handful of events.  The tau is  
monochromatic, and is opposite a monochromatic W and a light 
neutrino.  We know of no backgrounds to this signature, and a complete 
analysis would be worthwhile.
    
\section{The Two-Higgs Doublet and Randall-Sundrum Models}

\subsection{The Two-Higgs Doublet Model}            
    The minimal Standard Model Higgs sector consists of one complex
Higgs doublet.   One of the simplest and most popular
extensions of the Higgs sector is the two-Higgs doublet model (2HDM).   By 
requiring that all fermions of a given electric charge  couple to no more 
than one Higgs doublet \cite{gw}, one avoids flavor changing neutral 
currents.  This is accomplished with a simple $Z_2$ symmetry.
            
    The 2HDM is an attractive model for several reasons:
            \begin{itemize}
                \item it contains charged Higgs bosons and pseudoscalars
                \item it adds relatively few new arbitrary parameters
                \item it allows for spontaneous CP violation, and can give sufficient 
                baryogenesis
                \item this structure of the Higgs sector is required in low-energy supersymmetric 
                models
            \end{itemize}
A very detailed discussion of the 2HDM can be found in the Higgs Hunter's 
Guide \cite{hhg}.

    This model has two complex, $Y = 1$,
$SU(2)_L$ doublet scalar fields $\phi_1$ and $\phi_2$.   The vacuum 
expectation values of the neutral components of the Higgs doublets are 
$v_1$ and $v_2$, respectively.   It is useful to define
            \begin{equation}
                \tan \beta = \frac{v_2}{v_1}.
            \end{equation}
    The physical Higgs fields consist of two neutral scalars, a neutral 
pseudoscalar and a charged Higgs scalar.  In the charged sector, there will be both a Goldstone boson and a physical Higgs state.  The charged Higgs is given
 by
            \begin{equation}    
                H^\pm = - \phi_1^\pm \sin\beta + \phi_2^\pm \sin\beta
            \end{equation}
\begin{figure}[bth]
%\vspace*{-3.1in}
        \centerline{\epsfxsize 6.0in {\epsfbox{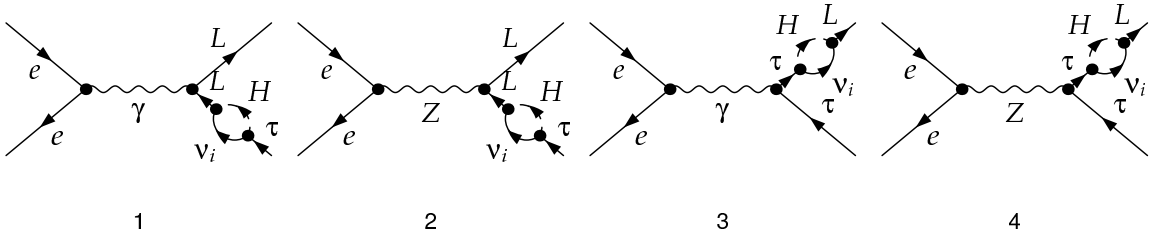}}}
%\vspace*{-2.35in}
\caption{The extra self energy diagrams contributing to $e^+e^-\to L\bar{\tau}$ in 2HDM.}
\label{fig:THDMself}
\end{figure}
\begin{figure}[bth]
\vspace*{0.2in}
        \centerline{\epsfxsize 6.0in {\epsfbox{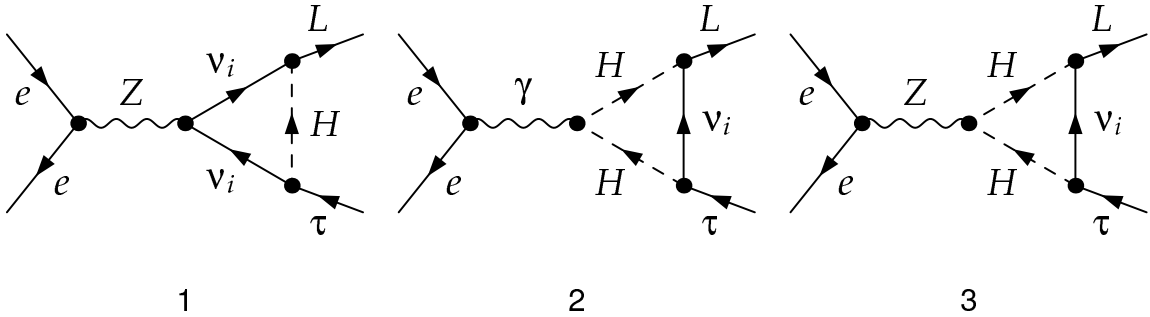}}}
%\vspace*{-2.35in}
\caption{The extra vertex diagrams contributing to $e^+e^-\to L\bar{\tau}$ in 2HDM.}
\label{fig:RHDMvertex}
\end{figure}
        
For our calculation, the neutral scalars will not contribute.  However, the 
charged Higgs boson will contribute.  One simply replaces the charged 
Goldstone boson $G$ in Figures 1-3  with the charged Higgs boson; 
these diagrams are shown in Figures 5 and 6.  The 
only exception is that the $Z W^\mp H^\pm$ vertex vanishes \cite{hhg}.  There 
are now two new parameters in the calculation, the mass of the charged 
Higgs boson and $\tan\beta$.    

There are two versions of the 2HDM.  In Model I, all of the fermions 
couple to one of the Higgs doublets; in Model II (which is included in 
supersymmetric models), the neutral leptons 
couple to one doublet and the charged leptons couple to the other.  
The relevant Yukawa couplings are
\begin{eqnarray}
L-N-H^+\; :&& \frac{i e}{2 \sqrt{2} m_W \sin\theta_W}\left[\frac{m_N}{\tan\beta}(1-\gamma_5) + m_L Y (1+\gamma_5)\right],\nonumber \\
e_i-N-H^+\; :&& \frac{-i e\delta_{i,3} V_{34}}{2 \sqrt{2} m_W \sin\theta_W}\left[\frac{m_N}{\tan\beta}(1-\gamma_5) + m_{ei} Y (1+\gamma_5)\right],\nonumber\\
L-\nu_i-H^+\; :&& \frac{i e \delta_{i,3} V_{34}}{2 \sqrt{2} m_W \sin\theta_W} m_L Y (1+\gamma_5)\,,
\end{eqnarray}
where $Y=-1/\tan\beta$ for Model I  and $\tan\beta$ for Model II and the vertices for the ordinary lepton - ordinary neutrino - $H^+$ can be 
found in the Higgs Hunter's Guide \cite{hhg}.
   
    Constraints from $b\rightarrow s\gamma$ force the mass of the charged Higgs 
to exceed approximately 200 GeV \cite{bsgamma}.   $\tan \beta$ and 
$\cot \beta$ must be less 
than about $3$ so that the charged and neutral lepton Yukawa
coupling remain perturbative.  
\begin{figure}[htb]
%\vskip -0.3in 
\begin{center}$
	\begin{array}{cc}
\hspace*{-1.4cm}
	\includegraphics[width=3.6in]{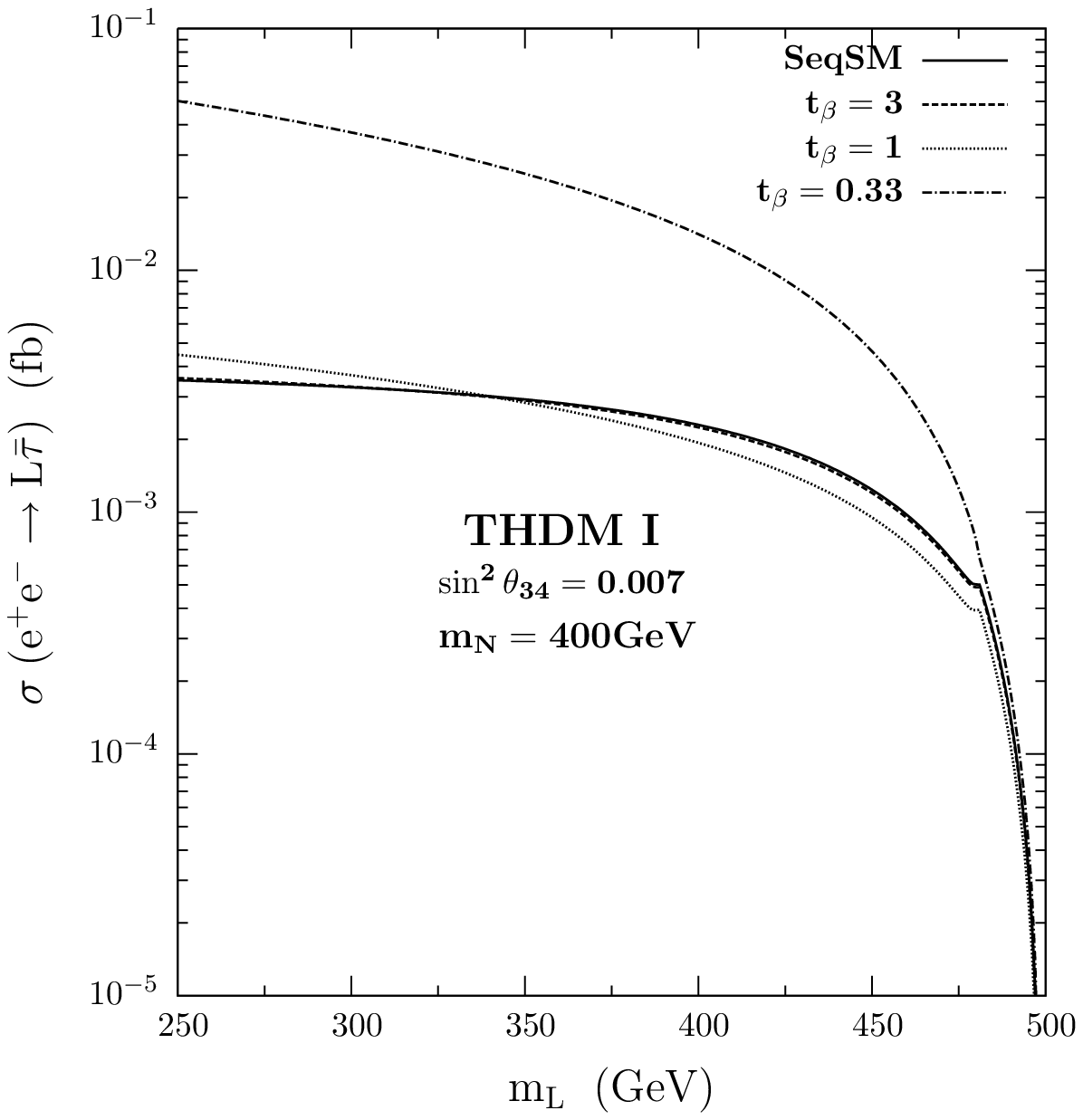} &\hspace*{-0.9cm}
	\includegraphics[width=3.6in]{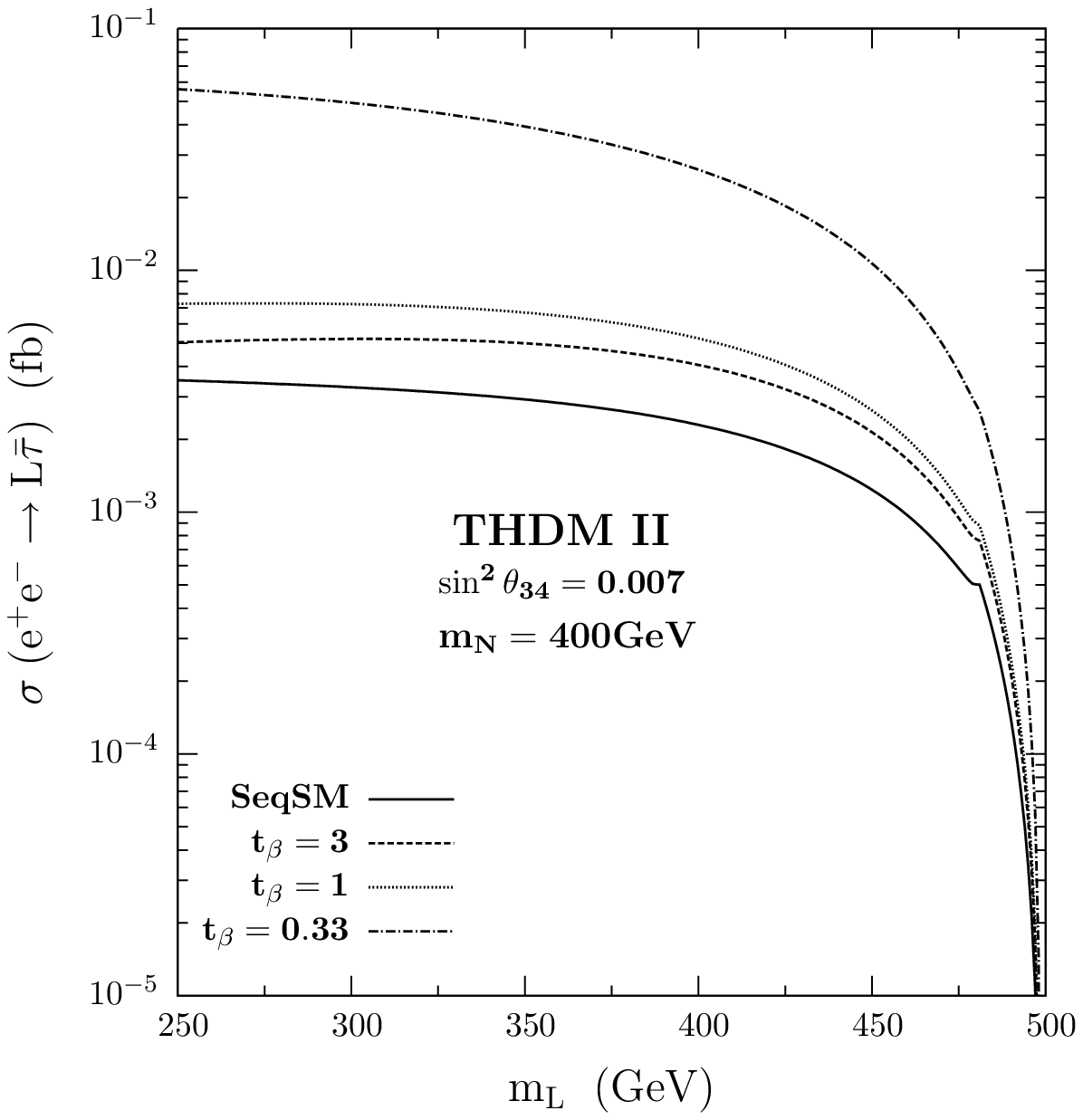}
	\end{array}$
\end{center}
\vskip -0.25in
      \caption{The total cross section of $e^+e^-\to L\bar{\tau}$ as a function of the heavy lepton mass $m_L$ for $\rm \sqrt{s}=500$ GeV in an unpolarized electron-positron beam for various $\tan\beta$ values in 2HDM. In both graphs, the heavy neutrino and charged Higgs masses are set 400 GeV and 200 GeV, respectively.}
\label{fig:THDM}
\end{figure}

The results are presented for Model I and for Model II in Figure \ref{fig:THDM}.   
We see small changes in the cross section for $1 < \tan\beta < 3$, but 
substantial changes for $1 < \cot\beta < 3$.  In both models, the 
cross section can be enhanced by up to a factor of ten, leading to 
much easier detection at the ILC.  Note that the vertices involving 
the heavy neutrino scales a $1/\tan\beta$, and thus the cross section 
is enhanced if $\tan\beta<1$ as seen from Figure \ref{fig:THDM}. 

\subsection{The Randall-Sundrum Model}
    Another popular extension of the Standard Model is the Randall-Sundrum 
(RS1) model \cite{rs}.   In this model a slice of $AdS_5$ space with curvature 
$k$ is compactified on an $S_1/Z_2$ orbifold with radius $R$.  The fixed 
points of the orbifold correspond to two 3-branes with opposite tensions.  
The 4D graviton is localized on the positive 
tension, or Planck brane, while the Higgs field is localized on the 
negative tension, or TeV brane.  This model generates a gauge hierarchy 
given by the warp factor, $e^{-\pi kR}$, and for $kR \simeq 12$ completely 
solves the hierarchy problem (see \cite{Frank:2007jb} for a bound on $kR$ from the Casimir force measurements).

    In the original formulation, the model had all of the fermions on the TeV 
brane.  More interesting phenomenology can occur when the Standard Model 
fermions and gauge bosons can propagate in the bulk \cite{bulk}.  In this 
case, the profiles of bulk fermion wavefunctions depend on their 5D mass parameters.  By 
choosing the lighter fermions to live near the Planck brane, one can 
naturally explain the small Yukawa couplings for the light fermions, since 
their overlap with the TeV-brane localized Higgs boson is exponentially 
supressed.  Thus the model can also explain the flavor hierarchy, since 
large differences in Yukawa couplings can arise from small differences in 
the mass parameters.  The flavor hierarchy simply becomes a matter of  
geography in the fifth dimension.

    In an interesting series of papers, Agashe, Perez and 
Soni \cite{a1,a2,a3} discussed the phenomenological implications of 
the flavor structure of these models.  They noted that one expects larger 
flavor changing neutral currents for the heavier generations, thus evading 
bounds involving light quarks.  In particular \cite{a3}, Agashe, et al.  considered top 
flavor violation at colliders, considering $t\rightarrow cZ$ at the LHC, 
and $e^+e^-\rightarrow t\bar{c}$ at the ILC.  Clearly, a similar process 
could lead to $e^+e^-\rightarrow L\tau$ as well, if a fourth generation 
exists.  The mechanism is caused by the fact that the couplings of the 
fermions to the gauge boson Kaluza-Klein (KK) modes are not universal due to 
the different profiles for the fermions, and mixing between the gauge KK 
modes and the gauge bosons leads to flavor violating couplings of the 
$Z$.  We refer the reader to Ref. \cite{a3} for details.

    One can simply carry over the calculation of $e^+e^-\rightarrow t\bar{c}$ 
in Ref. \cite{a3} to this model.  There is, however, one crucial 
difference.   Since the couplings of the left-handed $b$ quark to the $Z$ 
are measured to an accuracy of less than one percent and the $b$ quark is in 
a doublet with the left-handed top, one can not put the left-handed top and 
bottom too close to the TeV brane.  The right-handed top, however, can be 
close to the TeV brane.  Thus the top flavor violation is predominantly 
right-handed.  In the four generation case, there are no such restrictions, 
therefore the $L$ flavor violation is relatively unconstrained.  For 
definiteness, we choose the same magnitude for the left- and right-handed flavor 
violation, and set the coefficient of the 
$\bar{L}\gamma_\mu\tau Z^\mu$ term to be the same as that of the 
$\bar{t}_R\gamma_\mu c_R Z^\mu$ term in the Agashe, et al. analysis \cite{a3}.  This 
is not unreasonable, since $m_c/m_t \sim m_\tau/m_L$ indicates that 
similar mixing angles may be expected.

    Using this flavor violating coupling, one can find the total cross section 
for $e^+e^-\rightarrow L\tau$.  The result depends as well on the KK 
scale.  It has been shown \cite{a4} that a custodial $SU(2)$ symmetry in 
the bulk can allow the KK gauge boson mass to be as low as $3$ TeV, and 
perhaps somewhat lower if a modest fine-tuning is allowed, without 
conflicting with precision electroweak results.   Rather than 
calculate the interference with the Standard Model diagrams, we 
simply look at the RS model effects in isolation.  This is because the 
uncertainty in the flavor-violating couplings preclude precise 
calculations.   We find that if $M_{KK}$ 
is $1$ TeV, then the cross section varies from $1.0$ to $0.5$ femtobarns 
as the $L$ mass varies from $250$ GeV to $350$ GeV, and scales as 
$1/M_{KK}^4$.  Thus, we see a significant enhancement of the cross 
section in the KK mass range of $1-3$ TeV.   One should keep in mind 
that the KK gauge bosons, if they exist, will be discovered at the 
LHC long before the ILC is constructed.

\section{Detection and Conclusions}

    There are two possible decay modes for the $L$.  It can decay into $N W$, 
or into $\nu_\tau W$.  Of course,  if 
the $N$ is heavier only the latter decay is possible.   Regardless, there will be substantial missing 
energy in the decay.

    For the $e^+e^-\rightarrow L\tau$ process detection should be 
extremely straightforward, since the $\tau$ is monochromatic.  For an $L$ 
mass of $300$ GeV, the $\tau$ energy is 160 GeV, leading to a decay 
length, $\gamma ct$, of $0.8$ centimeters.   This is comparable to the size 
of the inner vertex detector at the ILC.   

    In a wide region of the 
mass-mixing angle plane, the $L$ will decay into a $\nu_\tau$ and a $W$. 
This would seem to give a clear signature, with a monochromatic $\tau$, a 
monochromatic $W$ and missing energy.  The primary background 
will be from $\tau$ pairs, where one of the $\tau$'s is misidentified.  A 
detailed Monte Carlo analysis is beyond the scope of this paper, but if 
the background can be eliminated, then a few events will suffice to 
discover the $L$.

    In the Standard Model case, we have found that there will be a 
    few events produced at the ILC, and the question of whether or 
    not the $L$ can be detected depends on the details of the 
    detector and Monte Carlo simulations.

    We then considered contributions from the charged Higgs boson of a 
2HDM, as well as flavor changing effects in the Randall-Sundrum model.  In 
both cases, there are regions of parameter space in which the cross 
section is substantially higher, leading to straightforward detection 
at the ILC.

Long before the ILC is built, the LHC will have determined whether or 
not a fourth generation exists.  If it does exist, then detection of 
the charged heavy lepton at the LHC will be very difficult and 
perhaps impossible.  At the ILC, if the mass of the heavy lepton is 
more than half $\sqrt{s}$, pair production will be impossible, and the 
process calculated in this paper may be the only mechanism for 
detection.

    We thank Chris Carone and Josh Erlich for their useful comments, 
    and David Reiss and Thomas Hahn for assistance with the software 
    packages.  This work is 
supported by the National Science Foundation, Grant PHY-0554854. The work of I.T. is supported by the NSERC of Canada under the Grant No. SAP01105354.

\end{document}